\def\mytuning#1{#1}         % This selects slightly larger one.
\documentstyle{article}

\mytuning{
 
 \textwidth 390pt
 \textheight 47.8\baselineskip
 }

\catcode`@=11
\def\@begintheorem#1#2{\it \trivlist
      \item[\hskip \labelsep{\bf #1\ #2:}] \rm}
\def\@opargbegintheorem#1#2#3{\it \trivlist
      \item[\hskip \labelsep{\bf #1\ #2\ (#3):}] \rm}
\def\incite#1{\if@filesw\immediate\write\@auxout{\string\citation{#1}}\fi
\def\@citea{}{\@for\@citeb:=#1\do
{\@citea\def\@citea{,\penalty\@m}\@ifundefined
{b@\@citeb}{{\bf ?}\@warning
{Citation `\@citeb' on page \thepage \space undefined}}%
\hbox{\csname b@\@citeb\endcsname}}}}
\catcode`@=12
\def\supcite#1{\raisebox{.8ex}{\scriptsize\incite{#1}}}

\def\thebibliography#1{
\section*{references}\list
{\arabic{enumi}}{\settowidth\labelwidth
{#1}\leftmargin\labelwidth
\advance\leftmargin\labelsep
\usecounter{enumi}}
\def\newblock{\hskip 1.1em plus .33em minus .07em}
\sloppy\clubpenalty4000\widowpenalty4000
\sfcode`\.=1000\relax}

\newtheorem{theorem}{Theorem}
\newtheorem{proposition}[theorem]{Proposition}
\newtheorem{definition}{Definition}
\newtheorem{assumption}{Assumption}

\font\EF=eufm10
\font\BI=cmbxti10

\newcommand{\spc}{\quad}
\newcommand{\rop}[1]{\!\! & \!\! #1 \!\! & \!\!}
\newcommand{\tallsp}{\raisebox{15pt}{}}

\newcommand{\complex}{\mbox{\BI C}}
\newcommand{\mgrC}{\complex\,\backslash\{0\}}
\newcommand{\ratint}{\mbox{\BI Z}}
\newcommand{\Cayley}{\mbox{\EF C}{}^C}
\newcommand{\univR}{{\cal R}}
\newcommand{\Gtwo}{\mbox{\rm G}_2}
\newcommand{\gee}{\mbox{\EF g}}

\newcommand{\pairing}[2]{\langle{#1},\:{#2}\rangle}
\newcommand{\dual}{^{\textstyle\ast}}
\newcommand{\qcombi}[3]{\Bigl[\mathop{}^{\textstyle\rule
 [-0.9ex]{0em}{1.2ex}#1}_{\textstyle #2}\Bigr]_{#3}}

\newcommand{\Fun}{\mbox{\rm Fun}}

\newcommand{\Mat}{\mbox{\rm Mat}}
\newcommand{\Aut}{\mbox{\rm Aut}\,}
\newcommand{\rank}{\mbox{\rm rank}\,}
\newcommand{\diag}{\mbox{\rm diag}\,}
\newcommand{\id}{\mbox{\rm id}}

\newcommand{\tgn }[2]{\,t^{#1}{}_{#2}}
\newcommand{\Rmat}[4]{\,R^{#1 #2}{}_{#3 #4}}
\newcommand{\CGg }[2]{\,g^{#1 #2}}
\newcommand{\CGgc}[2]{\,\tilde{g}_{#1 #2}}
\newcommand{\CGf }[3]{\,f^{#1 #2}{}_{#3}}
\newcommand{\CGfc}[3]{\,\tilde{f}{}^{#1}{}_{#2 #3}}
\newcommand{\Kdel}[2]{\,\delta^{#1}{}_{#2}}
\newcommand{\lgn }[3]{\,l^{(#1)#2}{}_{#3}}
\newcommand{\repr}[2]{\,\rho^{#1}{}_{#2}}
\newcommand{\Rtws}[5]{\,R^{(#1)#2 #3}{}_{#4 #5}}
\newcommand{\rep }[1]{\underline{#1}}
\newcommand{\crdn}[2]{\,w^{#1}{}_{#2}}

\newcommand{\wdsp}{\quad\quad\;\;}
\newcommand{\Rmx}[4]{\!R^{#1 #2}{}_{#3 #4}}
\newcommand{\SCg}[2]{\!g^{#1 #2}}
\newcommand{\Sgc}[2]{\!\tilde{g}_{#1 #2}}
\newcommand{\SCf}[3]{\!f^{#1 #2}{}_{#3}}
\newcommand{\Sfc}[3]{\!\tilde{f}{}^{#1}{}_{#2 #3}}

\begin{document}

\begin{flushright}
            Preprint OCU-150 (revised)
\end{flushright}
\bigskip
\begin{center}
 \LARGE     A Class of FRT Quantum Groups\\
            and $\Fun_q(\Gtwo)$ as a Special Case\bigskip\medskip\\
 \large     Norihito Sasaki\smallskip\\
 \normalsize
 \it        Department of Physics, Osaka City University,\\
            Sumiyoshi-ku, Osaka 558, Japan\medskip\\
 \rm        November 1993\\
            \makebox[0em][r]{revised:\,\,}December 1993
\end{center}

\mytuning{\vspace*{\fill}\vspace*{\fill}}
\begin{abstract}\normalsize
A class of quantum groups that includes a $\Gtwo$ type quantum group
$\Fun_q(\Gtwo)$ is considered.
After elucidating its connection with the quantized universal enveloping
algebras, a concrete construction of $\Fun_q(\Gtwo)$ is presented.
\end{abstract}
\mytuning{\vspace*{\fill}}

\section{introduction}

There are at least two approaches to quantum groups:
One\supcite{Dr,Ji,JiJ} is based on the Chevalley basis of Lie algebras and
gives, in terms of Hopf algebra, a remarkable deformation --- the quantized
universal enveloping (QUE) algebra --- for every Kac-Moody Lie algebra.
The other\supcite{FRT,RTF} --- known as Faddeev-Reshetikhin-Takhtajan (FRT)
approach --- is based on matrix relations that employ noncommuting matrix
entries and gives a quantization of Lie groups with retaining a formal
\mytuning{\pagebreak}
correspondence to the classical objects.

The former provides a way of constructing $R$-matrices, while the latter
needs an $R$-matrix as a starting point.  The latter, instead of this cost,
seems to have advantages in some directions, especially, in investigation
on noncommutative geometries\supcite{Wo} as pointed out in Ref.~\incite{Ju}.

Since there are $R$-matrices not explained currently in terms of the
QUE-algebras, the latter should give more various quantum groups apart
from the problem of incorporating an antipode.  Nevertheless, as far as
the author's knowledge, quantum groups of the exceptional type so far
have been constructed concretely only in the former approach.  So we
think it natural to ask how the exceptional Lie groups are
quantized in the former approach.  The preprint provides an answer in
part; we present a $\Gtwo$ type quantum group in the FRT approach after
giving a consideration on a class of quantum groups with slightly
general context and elucidating its connection with the QUE-algebras.

The work was inspired by Okubo's lecture note.\supcite{OkN}

\section{notations and conventions}
\label{sec-nc}

This section summarizes well-known notions and facts for
the definiteness on notations and conventions in the preprint.
For the concepts and the foundations of the FRT approach, we refer
the reader to Ref.~\incite{RTF}.

In the preprint an algebra means an algebra over the complex field
$\complex$.  Any algebra except for Lie algebras is supposed to have
the unity element~1.  We do not distinguish the unity element (if exist)
of an algebra from that of the base field $\complex\,$ unless otherwise
stated.

For any Hopf algebra $A$, the comultiplication, counit and antipode
are denoted by $\Delta$, $\epsilon$ and $S$ respectively.

For any Hopf algebra $A$, an $A$-module $V$ or representation
$(\rho_{V},\:V)$ of $A$ means merely a complex representation
as an algebra, however, the category of $A$-modules aquires extra
structures coming from $\Delta$, $\epsilon$ and $S$ for $A$:
For any two $A$-modules $V$ and $W$, there is the tensor product module
$V\otimes W$, which is a tensor product space endowed with the action
$\rho_{V\otimes W}:=(\rho_V \otimes \rho_W)\circ\Delta$.
There is the trivial module $\rep{1}$, which is a 1~dimensional
vector space endowed with the action $\rho_{\rep{1}}:=\epsilon$.
For any $A$-module $V$, there is the contragredient module $V\dual$,
which is a dual vector space endowed with the action $\rho_{V\dual}$
defined by
\[
\pairing{\rho_{V\dual}(x)\,\zeta}{v} = \pairing{\zeta}{\rho_V(S(x))\,v}
\spc\spc \forall x \in A,
\spc \forall \zeta \in V\dual,\spc \forall v \in V,
\]
where $\pairing{\;}{}$ stands for the pairing between $V\dual$ and $V$.
The symbol for the action is often suppressed in the formula such that
the module that the vectors appearing there belong to is clearly
identifiable.  The situation that $A$-modules $V$ and $W$ are isomorphic
is denoted by $V\simeq W$.

For any Hopf algebra $A$, the dual Hopf algebra $A\dual$ means a
dual space of $A$ endowed with the Hopf algebra structure defined by
\begin{eqnarray*}
&& (l\, l')(t) := (l \otimes l')(\Delta(t)),\spc
   1_{A\dual}(t) := \epsilon\,(t), \\
&& \Delta(l)(t \otimes t') := l\,(t\, t'),\spc
   \epsilon\,(l) := l\,(1_{A}),\spc
   S(l)(t) := l\,(S(t))
\end{eqnarray*}
with $l,\:l' \in A\dual$ and $t,\:t' \in A$.
Here $1_{A}$, $1_{A\dual}$ are the unities of $A$, $A\dual$ respectively,
and the pairing between the tensor space is defined by
$(l \otimes l')(t \otimes t') = l(t)\!\cdot l'(t')$ as usual.

A Hopf algebra $A$ is called quasitriangular\supcite{Dr} if there exists
an invertible element $\univR \in A \otimes A$ such that
\begin{eqnarray}
&& \Delta'(x) = \univR \Delta(x) \univR^{-1}\spc \forall x \in A,
  \label{eq:intwn}\\
&& (\Delta \otimes \id)(\univR) = \univR_{13} \univR_{23}, \label{eq:qtaa}\\
&& (\id \otimes \Delta)(\univR) = \univR_{13} \univR_{12}, \label{eq:qtab}
\end{eqnarray}
where $\Delta' := \sigma \circ \Delta$ is the opposite comultiplication and
$\univR_{12}:= \univR \otimes 1_A$, $\univR_{23}:= 1_A \otimes \univR$,
$\univR_{13}:= (\id \otimes \sigma)(\univR_{12})$.
In the above, $\sigma$ stands for the permutation on $A \otimes A$, namely,
$\sigma$:$\:x \otimes y \mapsto y \otimes x$.
Such an element $\univR$ is referred to as the universal $R$-matrix.
As a consequence of quasitriangularity, $\univR$ satisfies
\begin{eqnarray}
&(\epsilon\otimes\id)(\univR) = 1_A
 = (\id\otimes\epsilon)(\univR),&  \label{eq:RtoI} \\
&(S\otimes\id)(\univR) = \univR^{-1},
 \spc (\id\otimes S)(\univR^{-1}) = \univR.  \label{eq:RtoS}
\end{eqnarray}
More importantly, $\univR$ gives a solution to the Yang-Baxter equation
\begin{eqnarray}
&\univR_{12}\univR_{13}\univR_{23}=\univR_{23}\univR_{13}\univR_{12},&
  \label{eq:aYB}
\end{eqnarray}
which is an abstraction of the original form (\ref{eq:YB}).
A solution to (\ref{eq:YB}) is called an $R$-matrix.
A representation of the universal $R$-matrix gives an $R$-matrix.

The QUE-algebra $U_q(\gee)$ is a $q$-deformation of the universal enveloping
algebra of a Lie algebra $\gee$.  In the preprint it is assumed that $\gee$
is a finite dimensional simple Lie algebra and the deformation parameter
$q \in \mgrC$ is generic (not a root of unity).

Let $\alpha_i$ be the $i$\/th simple root for $\gee$.  The invariant inner
product $(\;,\:)$ in the root space defines the Cartan integers
$a_{ij}:= 2\,(\alpha_i,\:\alpha_j)/(\alpha_i,\:\alpha_i)$ and subsidiary
parameters $q_i := q^{(\alpha_i,\;\alpha_i)}$.  We follow the convention
of Ref.~\incite{JiJ} for the QUE-algebras.\supcite{Dr,Ji,JiJ}  The
QUE-algebra $U_q(\gee)$ is an associative algebra generated by
$X_i^{\pm}$, $K_i{}^{\pm 1}$ ($\: i = 1,\:\ldots\:,\:\rank\gee $)
satisfying the defining relations
\begin{eqnarray*}
&& K_i K_j = K_j K_i,\spc\spc K_i K_i{}^{-1} = K_i^{-1} K_i = 1, \\
&& K_i X_j^{\pm} K_i{}^{-1} = q_i{}^{\pm a_{ij}} X_j^{\pm}, \spc\spc
   X_i^{+} X_j^{-} - X_j^{-} X_i^{+} = \delta_{ij} \:
   \frac{K_i -  K_i{}^{-1}}{q_i - q_i{}^{-1}}, \\
&& \sum_{\nu = 0}^{1 - a_{ij}} (-1)^{\nu} \qcombi{1 - a_{ij}}{\nu}{q_i}
 (X_i^{\pm}){}^{1-a_{ij}-\nu} X_j^{\pm} \,(X_i^{\pm}){}^{\nu} = 0
 \spc\spc (i \neq j),
\end{eqnarray*}
where $\bigl[\raisebox{.15ex}{${}^{m}_{\,n}$}\bigr]_q$
stands for the $q$-binomial coefficient
\[
\qcombi{m}{n}{q} := \frac{q^{m} - q^{-m}}{q^{n} - q^{-n}} \cdot
 \frac{q^{m-1} - q^{1-m}}{q^{n-1} - q^{1-n}} \cdot \: \cdots \: \cdot
 \frac{q^{m-n+1} - q^{n-m-1}}{q - q^{-1}}
\]
defined for $n,\:m\in\ratint$, $\;0\leq n\leq m$.  The Hopf algebra structure
is introduced to the algebra $U_q(\gee)$ as follows:
\mytuning{\pagebreak}
\begin{eqnarray*}
&& \Delta(X_i^{+}) = X_i^{+} \otimes 1 + K_i \otimes X_i^{+}, \spc
   \Delta(X_i^{-}) = X_i^{-} \otimes K_i{}^{-1} + 1 \otimes X_i^{-},  \\
&& \Delta(K_i{}^{\pm 1}) = K_i{}^{\pm 1} \otimes K_i{}^{\pm 1}, \\
&& \epsilon\,(X_i^{\pm}) = 0,\spc
   \epsilon\,(K_i{}^{\pm 1}) = 1, \\
&& S(X_i^{+}) = -K_i{}^{-1} X_i^{+},\spc
   S(X_i^{-}) = -X_i^{-} K_i,\spc
   S(K_i{}^{\pm 1}) = K_i{}^{\mp 1}.
\end{eqnarray*}
The Hopf algebra $U_q(\gee)$ is known\supcite{Dr} to be quasitriangular.

It is easily seen by writing $K_i = q_i{}^{H_i}$ that the algebra in the
limit $q\rightarrow 1$ reduces to the universal enveloping algebra
$U(\gee)$; in the same limit $H_i$ and $X_i^{\pm}$ are identified as
generators of the Lie algebra $\gee$.

\section{a class of FRT quantum groups}
\label{sec-gen}

We consider a class of quantum groups that can be viewed as
missing ones in the FRT approach.\supcite{FRT,RTF}

We hereafter always suppose that latin indices $i,\:j,\:\ldots$ run over
the values $1,\:2,\:\ldots,\:N$ and that repeated indices imply summation
over them, where $N$ is a certain fixed integer (we will set $N=7$ for
the case of $\Fun_q(\Gtwo)$ in Sec.~\ref{sec-qG2}).

Let $A_{Rgf}$ be an associative algebra generated by
$\tgn{i}{j}$ satisfying the defining relations
\begin{eqnarray}
\Rmat{i}{j}{k}{l} \tgn{k}{m} \tgn{l}{n}
 \rop{=} \tgn{j}{l} \tgn{i}{k} \Rmat{k}{l}{m}{n}, \label{eq:RTT}\\
\CGg{i}{j} \rop{=} \tgn{j}{l} \tgn{i}{k} \CGg{k}{l}, \label{eq:TTG}\\
\CGgc{k}{l} \tgn{k}{m} \tgn{l}{n} \rop{=} \CGgc{m}{n}, \label{eq:TGC}\\
\CGf{i}{j}{k} \tgn{k}{m}
 \rop{=} \tgn{j}{l} \tgn{i}{k} \CGf{k}{l}{m}, \label{eq:TTF}\\
\CGfc{j}{k}{l} \tgn{k}{m} \tgn{l}{n}
 \rop{=} \tgn{j}{l} \CGfc{l}{m}{n}. \label{eq:TFC}
\end{eqnarray}
Here, $\Rmat{i}{j}{k}{l}$, $\CGg{i}{j}$, $\CGgc{i}{j}$, $\CGf{i}{j}{k}$
and $\CGfc{i}{j}{k}$ belong to $\complex\,$ and are referred to, in the
preprint, as the structure constants of $A_{Rgf}$.

We impose a condition on the structure constants in order to ensure that
the algebra $A_{Rgf}$ admits a Hopf algebra structure
\begin{eqnarray}
\Delta(\tgn{i}{j}) \rop{=} \tgn{i}{k} \otimes \tgn{k}{j}, \label{eq:Dt}\\
\epsilon\,(\tgn{i}{j}) \rop{=} \Kdel{i}{j}, \label{eq:ept}\\
S(\tgn{i}{j}) \rop{=} \CGg{l}{i} \tgn{k}{l} \CGgc{k}{j}, \label{eq:St}
\end{eqnarray}
that is to suppose
\begin{eqnarray}
& \CGg{i}{k} \CGgc{j}{k} = \Kdel{i}{j} = \CGg{k}{i} \CGgc{k}{j},
  & \label{eq:gs}\\
& \CGg{n}{i} \CGf{l}{m}{n} \CGgc{l}{j} \CGgc{m}{k}
 = \CGfc{i}{j}{k}
 = \CGg{i}{n} \CGf{l}{m}{n} \CGgc{j}{l} \CGgc{k}{m},
  & \label{eq:fs}\\
& \Rmat{i}{j}{k}{l} \CGg{k}{m} \CGg{l}{n}
 = \CGg{j}{l} \CGg{i}{k} \Rmat{m}{n}{k}{l}.
  & \label{eq:rg}
\end{eqnarray}
Using (\ref{eq:gs})--(\ref{eq:rg}), one can verify the Hopf algebra axioms
straightforwardly.

In addition to the conditions (\ref{eq:gs})--(\ref{eq:rg}), if
\begin{eqnarray}
\CGg{i}{j} \Kdel{k}{l}
 \rop{=} \Rmat{j}{k}{n}{p} \Rmat{i}{p}{m}{l} \CGg{m}{n}, \label{eq:RRg}\\
\Kdel{i}{j} \CGgc{k}{l}
 \rop{=} \CGgc{m}{n} \Rmat{m}{i}{k}{p} \Rmat{n}{p}{l}{j}, \\
\CGg{i}{j} \Kdel{k}{l}
 \rop{=} \Rmat{k}{i}{p}{m} \Rmat{p}{j}{l}{n} \CGg{m}{n}, \label{eq:RRgb}\\
\Kdel{i}{j} \CGgc{k}{l}
 \rop{=} \CGgc{m}{n} \Rmat{i}{n}{p}{l} \Rmat{p}{m}{j}{k}, \label{eq:gcRR}\\
\CGf{i}{j}{p} \Rmat{p}{k}{s}{l}
 \rop{=} \Rmat{j}{k}{n}{p} \Rmat{i}{p}{m}{l} \CGf{m}{n}{s}, \\
\Rmat{s}{i}{p}{j} \CGfc{p}{k}{l}
 \rop{=} \CGfc{s}{m}{n} \Rmat{m}{i}{k}{p} \Rmat{n}{p}{l}{j}, \\
\CGf{i}{j}{p} \Rmat{k}{p}{l}{s}
 \rop{=} \Rmat{k}{i}{p}{m} \Rmat{p}{j}{l}{n} \CGf{m}{n}{s}, \\
\Rmat{i}{s}{j}{p} \CGfc{p}{k}{l}
 \rop{=} \CGfc{s}{m}{n} \Rmat{i}{n}{p}{l} \Rmat{p}{m}{j}{k}  \label{eq:fcRR}
\end{eqnarray}
and the Yang-Baxter equation
\begin{eqnarray}
\Rmat{i_1}{i_2}{j_1}{j_2} \Rmat{j_1}{i_3}{k_1}{j_3}
 \Rmat{j_2}{j_3}{k_2}{k_3} \rop{=} \Rmat{i_2}{i_3}{j_2}{j_3}
 \Rmat{i_1}{j_3}{j_1}{k_3} \Rmat{j_1}{j_2}{k_1}{k_2} \label{eq:YB}
\end{eqnarray}
are satisfied, then the formula
\begin{eqnarray}
\lgn{\pm}{k_0}{k_\nu}(\tgn{i_1}{j_1}\cdots\tgn{i_\nu}{j_\nu})
 \rop{:=} \Rtws{\pm}{k_0}{i_1}{k_1}{j_1}\cdots
   \Rtws{\pm}{k_{\nu-1}}{i_\nu}{k_\nu}{j_\nu}  \label{eq:ldef}
\end{eqnarray}
(for $\nu=0$ in particular, $\lgn{\pm}{i}{j}(1) := \Kdel{i}{j}$) with
\begin{eqnarray}
\Rtws{+}{i}{j}{k}{l} \rop{:=} \Rmat{j}{i}{l}{k}, \nonumber \\
\Rtws{-}{i}{j}{k}{l} \rop{:=} R^{-1\;ij}{}_{kl}
 \spc \mbox{(inverse as a matrix)}    \label{eq:Rtw} \\
 \rop{=} \CGg{m}{i} \Rmat{n}{j}{m}{l} \CGgc{n}{k}
 = \CGg{j}{m} \Rmat{i}{n}{k}{m} \CGgc{l}{n} \nonumber
\end{eqnarray}
is well-defined as a defintion of the linear mappings
$\lgn{\pm}{i}{j}$: $A_{Rgf} \rightarrow \complex$.

The linear mappings $\lgn{\pm}{i}{j}$ by definition can be considered
to be in the dual Hopf algebra $A_{Rgf}\dual$.
A tedious calculation leads to the relations in $A_{Rgf}\dual$
\begin{eqnarray}
\Rmat{i}{j}{k}{l} \lgn{\pm}{l}{n} \lgn{\pm}{k}{m}
 \rop{=} \lgn{\pm}{i}{k} \lgn{\pm}{j}{l} \Rmat{k}{l}{m}{n}, \label{eq:dRTT}\\
\Rmat{i}{j}{k}{l} \lgn{+}{l}{n} \lgn{-}{k}{m}
 \rop{=} \lgn{-}{i}{k} \lgn{+}{j}{l} \Rmat{k}{l}{m}{n}, \label{eq:cRTT}\\
\CGg{i}{j} \rop{=} \lgn{\pm}{i}{k} \lgn{\pm}{j}{l} \CGg{k}{l}, \\
\CGgc{k}{l} \lgn{\pm}{l}{n} \lgn{\pm}{k}{m} \rop{=} \CGgc{m}{n}, \\
\CGf{i}{j}{k} \lgn{\pm}{k}{m}
 \rop{=} \lgn{\pm}{i}{k} \lgn{\pm}{j}{l} \CGf{k}{l}{m}, \\
\CGfc{j}{k}{l} \lgn{\pm}{l}{n} \lgn{\pm}{k}{m}
 \rop{=} \lgn{\pm}{j}{l} \CGfc{l}{m}{n}
\end{eqnarray}
and also leads to the expressions for $\Delta$, $\epsilon$ and
$S$ for $\lgn{\pm}{i}{j} \in A_{Rgf}\dual$
\begin{eqnarray}
\Delta(\lgn{\pm}{i}{j}) \rop{=} \lgn{\pm}{i}{k}\otimes\lgn{\pm}{k}{j}, \\
\epsilon\,(\lgn{\pm}{i}{j}) \rop{=} \Kdel{i}{j}, \\
S(\lgn{\pm}{i}{j}) \rop{=} \CGg{i}{l} \lgn{\pm}{k}{l} \CGgc{j}{k}.
\end{eqnarray}

We would like to mention that these equations excluding those employing
$\CGf{i}{j}{k},\:\CGfc{i}{j}{k}$ have been considered in
Ref.~\incite{RTF}\ for the construction of B$_{n}$, C$_{n}$
and D$_{n}$ type quantum groups.  Especially, the set of formulae
(\ref{eq:RTT}) and (\ref{eq:YB})--(\ref{eq:cRTT})
is inherent in the FRT approach.\supcite{FRT,RTF}

\section{connection with the QUE-algebras}
\label{sec-der}

We will look for the structure constants of a quantum group
considered in the preceding section in a certain representation of
a QUE-algebra\supcite{Dr,Ji,JiJ} of Drinfel'd-Jimbo.
Equations (\ref{eq:RTT}) and (\ref{eq:YB}) --- main relations
in the FRT approach\supcite{FRT,RTF} --- were the very origin of the
QUE-algebras; a connection between these relations and the
representations of the QUE-algebras was already noticed in the
pioneering work\supcite{Dr} of Drinfel'd (see Ref.~\incite{Bur}\ also).
The following proposition describes this well-known fact:
\begin{proposition}
\label{wellkn}
Let $U_q(\gee)$ be a QUE-algebra and $\univR $ be the universal
$R$-matrix.  Suppose that $V$ is a $U_q(\gee)$-module, $\dim V = N$.
And let us define $\Rmat{i}{j}{k}{l}\in\complex\,$ and
$\repr{i}{j}\in U_q(\gee)\dual$ by
\begin{eqnarray}
\univR \: v_k \otimes v_l \rop{=}
  v_i \otimes v_j \Rmat{i}{j}{k}{l}, \label{eq:RcR}\\
 x \: v_j \rop{=} v_i \repr{i}{j}(x)
  \spc\spc \forall x \in U_q(\gee)  \label{eq:matr}
\end{eqnarray}
with $\{v_1,\:\ldots,\:v_N\}$ being a basis of $V$.
Then $\Rmat{i}{j}{k}{l}$ gives a solution to the Yang-Baxter
equation (\ref{eq:YB}).
Reading $\tgn{i}{j}$ in the preceding section as $\repr{i}{j}$,
one finds (\ref{eq:RTT}), (\ref{eq:Dt}) and (\ref{eq:ept})
as equations for the dual Hopf algebra $U_q(\gee)\dual$.
\end{proposition}
\begin{proof}
The Yang-Baxter equation is obtained as a representation of (\ref{eq:aYB}).
The others are translations of $U_q(\gee)$'s some properties
into the language of the dual Hopf algebra $U_q(\gee)\dual$:
The relation (\ref{eq:RTT}) follows from (\ref{eq:intwn}).  The formula
(\ref{eq:Dt}) together with (\ref{eq:ept}) merely says that
$U_q(\gee)\ni x \mapsto (\repr{i}{j}) \in \Mat(N,\,\complex\,)$ is an algebra
homomorphism ({\it i.e.}, that $V$ is a $U_q(\gee)$-module).
\end{proof}

The setting in the proposition without any more restriction seems
inadequate to give a closed expression for $S(\repr{i}{j})$, however,
\[
S(\repr{i}{k})\,\repr{k}{j} = \Kdel{i}{j} = \repr{i}{k}\,S(\repr{k}{j})
\]
is satisfied quite generally due to the Hopf algebra axioms.

Just like in the proposition, we assume without notice that $\tgn{i}{j}$
is read as $\repr{i}{j}$ whenever equations in the preceding section are
referred in this section.  In addition to the setting in the proposition,
we from now on assume:
\begin{assumption}
\label{assump}
The $U_q(\gee)$-module $V$ is an $N$ (finite) dimensional vector space
over $\complex\,$ and its square has the irreducible decomposition
\[
  V \otimes V \simeq \rep{1} \oplus V \oplus \cdots,
\]
where we mean the right-hand side for expressing appearances of
$\rep{1}$ and $V$ with their multiplicity~1 and only~1.
\end{assumption}
We stress the following:
The assumption includes that $V$ is irreducible.
Since any finite dimensional $U_q(\gee)$-module is known\supcite{Ro} to
\mytuning{\pagebreak}
be completely reducible if $q$ is generic, the modules $V\dual$ and
$V \otimes V$ in particular are completely reducible
($q$ have been assumed generic).

The following definition of $\CGg{i}{j}$ and $\CGf{i}{j}{k}$
makes sense under the assumption:
\begin{definition}
\label{definn}
Let $\rep{1}'$ and $V'$ be the irreducible submodules of $V \otimes V$
such that
\[
V \otimes V = \rep{1}' \oplus V' \oplus \cdots,\spc\spc
 \rep{1}' \simeq \rep{1},\spc V' \simeq V.
\]
Define $\CGg{i}{j}$ and $\CGf{i}{j}{k}$ as Clebsch-Gordan coefficients
as follows.
\begin{eqnarray}
\rep{1}' \ni v'{}_0 \rop{=} v_j \otimes v_i \CGg{i}{j},  \\
V' \ni v'{}_k \rop{=} v_j \otimes v_i \CGf{i}{j}{k},
\end{eqnarray}
where $\{v'{}_0\}$ is a basis of the submodule $\rep{1}'$, namely,
\begin{eqnarray}
&&\Delta(x)\, v'{}_0 = v'{}_0 \,\epsilon\,(x) \spc \forall x \in U_q(\gee)
\end{eqnarray}
and $\{v'{}_1,\:\ldots,\:v'{}_N\}$ is a basis of the submodule $V'$
such that $v'{}_j \mapsto v_j$ defines a $U_q(\gee)$-isomorphism
$V'\rightarrow V$, namely,
\begin{eqnarray}
&&\Delta(x)\, v'{}_j = v'{}_i \repr{i}{j}(x) \spc \forall x \in U_q(\gee)
 \label{eq:Vpr}
\end{eqnarray}
with $\repr{i}{j}$ appeared in (\ref{eq:matr}).
\end{definition}
Then one readily finds (\ref{eq:TTG}) and (\ref{eq:TTF}).
\begin{proposition}
\label{isop}
There is a $U_q(\gee)$-isomorphism
$\varphi^{(1)}$: $V\dual\rightarrow V$,
\begin{eqnarray}
&\varphi^{(1)}(\zeta^j) = v_i \CGg{i}{j},&
\end{eqnarray}
where $\{\zeta^1,\:\ldots,\:\zeta^N\}$ is the dual basis of
$\{v_1,\:\ldots,\:v_N\}$, {\it i.e.}, $\pairing{\zeta^i}{v_j}=\Kdel{i}{j}$.
\end{proposition}
\begin{proof}
In terms of $\zeta^i\in V\dual$, the contragredient representation
is expressed as
\begin{eqnarray}
&& x \,\zeta^i = \zeta^j\,S(\repr{i}{j})(x)
 \spc \forall x \in U_q(\gee).  \label{eq:crep}
\end{eqnarray}
Using relation (\ref{eq:TTG}), we calculate
\begin{eqnarray*}
\varphi^{(1)}(x \,\zeta^k)
\rop{=} \varphi^{(1)}(\zeta^j)\,S(\repr{k}{j})(x)
 = v_i \CGg{i}{j}\,S(\repr{k}{j})(x)) \\
\rop{=} v_i \bigl(S(\repr{k}{j})\CGg{i}{j}\bigl)(x)
 = v_i \bigl(S(\repr{k}{j})\repr{j}{l}\repr{i}{m}\CGg{m}{l}\bigl)(x) \\
\rop{=} v_i \repr{i}{m}(x) \CGg{m}{k}
 = x \,v_m \CGg{m}{k} = x \,\varphi^{(1)}(\zeta^k)
\end{eqnarray*}
for $\forall x \in U_q(\gee)$, hence $\varphi^{(1)}$
is a $U_q(\gee)$-homomorphism.  Moreover, $\varphi^{(1)}$ is a nonzero
$U_q(\gee)$-homomorphism because $v_0\neq 0$ and so is the matrix
$(\CGg{i}{j})$.  Since $V$ is irreducible and $V\dual$ is completely
reducible, we see that Schur's lemma asserts existence of a submodule W
such that $W\subset V\dual$, $\;W\simeq V$.
Comparing the dimensions of these modules, we find $V\dual\simeq V$.
Therefore $\varphi^{(1)}$ is a $U_q(\gee)$-isomorphism.
\end{proof}
\begin{definition}
\label{defnb}
Since $V\dual$ is isomorphic to $V$, we have
\[
V\dual\otimes V\dual = \rep{1}^{\sim} \oplus V^{\sim} \oplus \cdots,
 \spc\spc \rep{1}^{\sim} \simeq \rep{1},
 \spc V^{\sim} \simeq V\dual\:(\simeq V).
\]
Here, as a consequence of the proposition, the submodules isomorphic
to $\rep{1}$ or $V$ are exhausted by those appeared above.
So we define $\CGgc{i}{j}$ and $\CGfc{i}{j}{k}$ by
\begin{eqnarray}
\rep{1}^{\sim} \ni \tilde{\zeta}{}^0
 \rop{=} \CGgc{i}{j} \,\zeta^j \otimes \zeta^i,  \\
V^{\sim} \ni \tilde{\zeta}{}^k
 \rop{=} \CGfc{k}{i}{j} \,\zeta^j \otimes \zeta^i,
\end{eqnarray}
where $\{\tilde{\zeta}{}^0\}$ is a basis of the submodule
$\rep{1}^{\sim}$, namely,
\begin{eqnarray}
&&\Delta(x)\, \tilde{\zeta}{}^0 = \tilde{\zeta}{}^0 \,\epsilon\,(x)
 \spc \forall x \in U_q(\gee)
\end{eqnarray}
and $\{\tilde{\zeta}{}^1,\:\ldots,\:\tilde{\zeta}{}^N\}$ is a basis of
the submodule $V^{\sim}$ such that $\tilde{\zeta}{}^j \mapsto \zeta^j$
defines a $U_q(\gee)$-isomorphism $V^{\sim}\rightarrow V\dual$, namely,
\begin{eqnarray}
&&\Delta(x)\, \tilde{\zeta}{}^i = \tilde{\zeta}{}^j\,S(\repr{i}{j})(x)
 \spc \forall x \in U_q(\gee)  \label{eq:vtil}
\end{eqnarray}
with $S(\repr{i}{j})$ appeared in (\ref{eq:crep}).
\end{definition}
Using the inverse of $S$, one readily finds (\ref{eq:TGC}) and
(\ref{eq:TFC}).
\begin{proposition}
There is a $U_q(\gee)$-isomorphism
$\varphi^{(-1)}$: $V\rightarrow V\dual$,
\begin{eqnarray}
&\varphi^{(-1)}(v_i) = \CGgc{i}{j} \,\zeta^j,&
\end{eqnarray}
which, apart from a scalar factor, gives the inverse of $\varphi^{(1)}$.
\end{proposition}
\begin{proof}
Most of the proof goes parallel to the preceding one.  The last statement
follows from irreducibility of $V$ and Schur's lemma.
\end{proof}
Now we adjust the scalar factor to fulfill
\begin{eqnarray}
&\varphi^{(1)}\circ \varphi^{(-1)} = \id_V,\spc
 \varphi^{(-1)}\circ \varphi^{(1)} = \id_{V\dual},& \label{eq:adsta}
\end{eqnarray}
by using scaling ambiguity in the definition of $v'{}_0$ or
$\tilde{\zeta}{}^0$.
Then (\ref{eq:gs}) is apparent and (\ref{eq:St}) is easily obtained.

We have already verified (\ref{eq:RTT})--(\ref{eq:gs}) and (\ref{eq:YB}).

\begin{proposition}
\label{newprop}
There exists a $U_q(\gee)$-isomorphism $\varphi$:
 $\,(V\otimes V)\dual \rightarrow V\dual\otimes V\dual$,
\begin{eqnarray}
& \varphi(\zeta^i\otimes\zeta^j)
 := \Rmat{i}{j}{k}{l}\,\zeta^k\otimes\zeta^l.&
\end{eqnarray}
\end{proposition}
\begin{proof}
We think it appropriate here to use the symbol for the action to a module.
The modules $(V\otimes V)\dual$ and $V\dual\otimes V\dual$
are identical as vector spaces, but the actions are different;
\begin{eqnarray}
\rho_{V\dual\otimes V\dual}
 \rop{=} (\rho_{V\dual}\otimes\rho_{V\dual})\circ\Delta,  \label{eq:actna}\\
\rho_{(V\otimes V)\dual}
 \rop{=} (\rho_{V\dual}\otimes\rho_{V\dual})\circ\Delta'. \label{eq:actnb}
\end{eqnarray}
The former (\ref{eq:actna}) is by defintion.  The latter (\ref{eq:actnb})
is due to coalgebra anti-homomorphism property of an antipode;
$\Delta\circ S = (S\otimes S)\circ\Delta'$.  However, as is well-known,
\mytuning{\pagebreak}
the both modules are isomorphic because (\ref{eq:intwn}) gives an
isomrphism.  Using (\ref{eq:RtoS}) and (\ref{eq:RcR}), we verify that
$\varphi$ is the isomorphism.
\end{proof}
Let $\{\zeta'{}^1,\:\ldots,\:\zeta'{}^N\}$ be the dual basis of
$\{v'{}_1,\:\ldots,\:v'{}_N\}$, {\it i.e.},
$\pairing{\zeta'{}^i}{v'{}_j}=\Kdel{i}{j}$.  Then
\begin{eqnarray}
&& \Delta'(x) \,\zeta'{}^i = \zeta'{}^j\,S(\repr{i}{j})(x)
 \spc \forall x \in U_q(\gee) \label{eq:Vprd}
\end{eqnarray}
follows from (\ref{eq:Vpr}) on the understanding that $\zeta'{}^i$
belongs to the contragredient module $(V')\dual$.  Note that use of
$\Delta'$ or $\rho_{(V\otimes V)\dual}$ is relevant in the above
because $(V')\dual$ is a submodule of $(V\otimes V)\dual$.

Since $(\rep{1}')\dual \simeq \rep{1}$ and $(V')\dual \simeq V$
(the former is easy and not so crucial; the latter follows
from proposition~\ref{isop}), we have
\[
 (V\otimes V)\dual = (\rep{1}')\dual \oplus (V')\dual \oplus \cdots,
 \spc\spc  (\rep{1}')\dual \simeq \rep{1},
 \spc  (V')\dual \simeq V\dual\:(\simeq V).
\]
Here, as a consequence of proposition~\ref{newprop}, the irreducible
submodules isomorphic to $\rep{1}$ or $V$ are exhausted by those appeared
above.  Therefore, Schur's lemma uniquely identify $\zeta'{}^i$ as
\begin{eqnarray}
\zeta'{}^i = c^{(0)}\CGfc{i}{j}{k}\,\zeta^j\otimes\zeta^k  \label{eq:upbt}
\end{eqnarray}
up to the constant $c^{(0)}\in\mgrC$; for, comparing with
(\ref{eq:vtil}), one immediately verify (\ref{eq:Vprd}) with (\ref{eq:upbt})
substituted.  So we arrive at
\begin{proposition}
There is an identity
\begin{eqnarray}
&& \CGfc{i}{k}{l}\CGf{l}{k}{j} = \frac{1}{c^{(0)}}\Kdel{i}{j}
 \spc \exists c^{(0)}\in\mgrC.  \label{eq:FFsId}
\end{eqnarray}
\end{proposition}
\begin{proof}
This is shown by evaluating $\pairing{\zeta'{}^i}{v'{}_j}$.
\end{proof}
\begin{proposition}
The linear mappings $\chi^{(\nu)}$:
$V^{\sim}\rightarrow V\dual\otimes V\dual$, $\:\nu = 1,\:2$ defined by
\begin{eqnarray}
 \chi^{(1)}(\tilde{\zeta}{}^j) \rop{=}
  \CGg{i}{j}\CGf{k}{l}{i}\CGgc{k}{m}\CGgc{l}{n}\,\zeta^n\otimes\zeta^m, \\
 \chi^{(2)}(\tilde{\zeta}{}^j) \rop{=}
  \CGg{j}{i}\CGf{k}{l}{i}\CGgc{m}{k}\CGgc{n}{l}\,\zeta^n\otimes\zeta^m
\end{eqnarray}
are shown to be
\begin{eqnarray}
&& \chi^{(1)} = \chi^{(2)} = c\,\id_{V^{\!\sim}}
 \spc \exists c \in\mgrC,
\end{eqnarray}
where, strictly speaking, we mean $\id_{V^{\!\sim}}$ as the inclusion mapping
$V^{\sim}\rightarrow V\dual\otimes V\dual$.
\end{proposition}
\begin{proof}
Similarly to the calculation for $\varphi^{(1)}$, commutativity of
$\chi^{(\nu)}$ with the action of $\forall x \in U_q(\gee)$ can be
verified (for $\nu=2$, the verification will be made easier by writing
$x=S^{-1}(y)$).  Hence these mappings are $U_q(\gee)$-homomorphisms.
Moreover, these mappings are nonzero $U_q(\gee)$-homomorphisms
because $(\CGg{i}{j})$ and $(\CGgc{i}{j})$ are regular as matrices
and some of the constants $\CGf{i}{j}{k}$ by definition are nonzero.
Therefore Schur's lemma with the irreducible decomposition of
$V\dual\otimes V\dual$ states that
\[
\chi^{(\nu)} = c^{(\nu)}\,\id_{V^{\sim}}
 \spc \exists c^{(\nu)} \in\mgrC.
\]
This says
\begin{eqnarray}
\CGg{n}{i}\CGf{l}{m}{n}\CGgc{l}{j}\CGgc{m}{k} \rop{=}
  c^{(1)}\CGfc{i}{j}{k},  \label{eq:fafc}\\
\CGg{i}{n}\CGf{l}{m}{n}\CGgc{j}{l}\CGgc{k}{m} \rop{=}
  c^{(2)}\CGfc{i}{j}{k}.  \label{eq:fbfc}
\end{eqnarray}
Here, $c^{(1)}=c^{(2)}$ because
\begin{eqnarray*}
\frac{c^{(1)}}{c^{(0)}}\Kdel{i}{j}
\rop{=} c^{(1)}\CGfc{i}{k}{l}\CGf{l}{k}{j}
 = \CGg{n}{i}\CGf{p}{m}{n}\CGgc{p}{k}\CGgc{m}{l}\CGf{l}{k}{j} \\
\rop{=} \CGg{n}{i}(\CGgc{p}{k}\CGgc{m}{l}\CGf{l}{k}{j})\CGf{p}{m}{n}
 = \CGg{n}{i}(c^{(2)}\CGfc{k}{m}{p}\CGgc{k}{j})\CGf{p}{m}{n} \\
\rop{=} \frac{c^{(2)}}{c^{(0)}}\Kdel{k}{n}\CGg{n}{i}\CGgc{k}{j}
 = \frac{c^{(2)}}{c^{(0)}}\Kdel{i}{j}.
\end{eqnarray*}
So we get the proposition by putting $c=c^{(1)}=c^{(2)}$.
\end{proof}
The value of $c$ is a matter of convension;
using scaling ambiguity in the definition of $v'{}_j$ or
$\tilde{\zeta}{}^j$, we can adjust
\begin{eqnarray}
& c\,\, (= c^{(1)}=c^{(2)}) = 1.& \label{eq:adstb}
\end{eqnarray}
Then (\ref{eq:fafc}) and (\ref{eq:fbfc}) turn out to be (\ref{eq:fs}).

As is easily verified, (\ref{eq:rg}) follows from (\ref{eq:RRg}) and
(\ref{eq:RRgb}).  So (\ref{eq:RRg})--(\ref{eq:fcRR}) are at this stage
all remainnig equations that we wish to show.  These rely mostly on
quasitriangularity of $U_q(\gee)$.
The formulae (\ref{eq:qtaa}) and (\ref{eq:qtab}) are represented
on the module $V\otimes V\otimes V$ as follows:
\begin{eqnarray}
&& (\repr{i_1}{j_1}\repr{i_2}{j_2}\otimes\repr{i_3}{j_3})(\univR)
  = \Rmat{i_1}{i_3}{j_1}{k_3}\Rmat{i_2}{k_3}{j_2}{j_3}, \label{eq:qtRa}\\
&& (\repr{i_1}{j_1}\otimes\repr{i_2}{j_2}\repr{i_3}{j_3})(\univR)
  = \Rmat{i_1}{i_3}{k_1}{j_3}\Rmat{k_1}{i_2}{j_1}{j_2}. \label{eq:qtRb}
\end{eqnarray}
Multiplying (\ref{eq:qtRa}) by $\CGg{j_2}{j_1}$, $\!\CGgc{i_1}{i_2}$,
$\!\CGf{j_2}{j_1}{j}\,$ or $\!\CGfc{i}{i_1}{i_2}$, we get four formulae.
Multiplying (\ref{eq:qtRb}) by $\CGg{j_3}{j_2}$, $\!\CGgc{i_2}{i_3}$,
$\!\CGf{j_3}{j_2}{j}\,$ or $\!\CGfc{i}{i_2}{i_3}$, we get four more formulae.
Using (\ref{eq:TTG})--(\ref{eq:TFC}) and (\ref{eq:RtoI}) suitably,
we can turn these eight formulae into (\ref{eq:RRg})--(\ref{eq:fcRR}).
This derivation with slightly general setting is found in Ref.~\incite{Res}.
One can derive (\ref{eq:RRg})--(\ref{eq:gcRR}) from (\ref{eq:RtoS}) also.

Although definition~\ref{defnb} with (\ref{eq:adsta}) and
(\ref{eq:adstb}) gives $\CGgc{i}{j}$ and $\CGfc{i}{j}{k}$,
it is perhaps more convenient for the purpose of computing these
constants to use (\ref{eq:gs}) and (\ref{eq:fs}) instead.
In this point of view, we summarize the arguments above as follows:
\begin{theorem}
\label{theor}
Suppose that a $U_q(\gee)$-module $V$ fulfills the assumption~\ref{assump}.
Let $(\Rmat{i}{j}{k}{l})$ be the $R$-matrix defined by (\ref{eq:RcR}) and
let $\CGg{i}{j}$ and $\CGf{i}{j}{k}$ be the constants given in
definition~\ref{definn}.  The remaining constants $\CGgc{i}{j}$ and
$\CGfc{i}{j}{k}$ are given by the 1st (or equivalently the 2nd) equations
in (\ref{eq:gs}) and (\ref{eq:fs}) respectively.  Then these constants
satisfy all of the equations (\ref{eq:gs})--(\ref{eq:YB}).

Furthermore, $\repr{i}{j}\in U_q(\gee)\dual$ defined by (\ref{eq:matr})
satisfies (\ref{eq:RTT})--(\ref{eq:St}) with these constants.
In other words, $\tgn{i}{j}\mapsto\repr{i}{j}$ defines a Hopf algebra
homomorphism $A_{Rgf}\rightarrow U_q(\gee)\dual$, where $A_{Rgf}$
is a Hopf algebra considered in the preceding section with these
constants as the structure constants.
\end{theorem}

\section{construction of $\Fun_q(\Gtwo)$}
\label{sec-qG2}

We notice that the $7$ dimensional irreducible module of the $\Gtwo$
type QUE-algebra $U_q(\gee_2)$ fulfills the assumption~\ref{assump};
the irreducible decomposition
\begin{eqnarray}
& \rep{7} \otimes \rep{7} \simeq \rep{1} \oplus \rep{7}
   \oplus \rep{14} \oplus \rep{27}&   \label{eq:svnsvn}
\end{eqnarray}
is valid for generic $q$, where $\rep{n}$ stands for the $n$ dimensional
irreducible $U_q(\gee_2)$-module.  Therefore this representation
induces a Hopf algebra $A_{Rgf}$ considered in Sec.~\ref{sec-gen} via
the prescription for the structure constants described in the
theorem~\ref{theor}.  This Hopf algebra is denoted by $A(q)$ temporarily.

Let us consider the situation at $q=1$:
\begin{proposition}
Suppose that $(\crdn{i}{j}) \in \Mat(7,\,\complex\,)$ is a solution
to the equations (\ref{eq:RTT})--(\ref{eq:TFC}) with the structure
constants of $A(1)$ provided that one reads $\tgn{i}{j}$ in the
equations as $\crdn{i}{j}$.
Let $G$ be the set of all such solutions $(\crdn{i}{j})$ and
define a multiplication between two elements of $G$ as matrices.
Then $G$ constitutes a Lie group isomorphic to $\Aut\Cayley$,
where $\Cayley$ stands for Cayley algebra over $\complex$.
\end{proposition}

Before going into the proof, let us remember the fact that
$\Aut\Cayley$ defines a $\Gtwo$ type complex Lie group.
It is known\supcite{Yo} that $\Aut\Cayley$ is simply-connected
and has the center of order~1.
Theory of Lie groups\supcite{Ise} hence deduces that any connected
complex Lie group of $\Gtwo$ type is isomorphic to $\Aut\Cayley$.
\begin{proof}
It is easily shown that $G$ constitute a group; the verification goes
parallel to that of the Hopf algebra axioms for $A_{Rgf}$.

We first consider the relation
\[
\zeta^j \,\zeta^i = -\mu\CGg{i}{j} + \CGf{i}{j}{k}\,\zeta^k, \spc
\mu = \frac{\CGfc{j}{k}{l} \CGf{l}{k}{j}}{42},
\]
where the constants $\CGg{i}{j}$, $\CGf{i}{j}{k}$ and $\CGfc{i}{j}{k}$
are those for $q=1$ also.
It is known\supcite{OkP,OkN} that the relation defines Cayley
algebra $\Cayley $ with the basis $\{1,\:\zeta^1,\:\ldots,\:\zeta^7\}$.
As is readily verified, for arbitrary $(\crdn{i}{j}) \in G$
the transformation $\zeta^i \mapsto \crdn{i}{j} \,\zeta^j$
preserves the relation above, therefore $G \subset \Aut\Cayley$.

Next we use the fact that $\rep{7}$ is a faithful representation
of the $\Gtwo$ type Lie algebra $\gee_2$.  This fact can be seen
with an appearance of the adjoint representation $\rep{14}$ (the
adjiont representation of a semi-simple Lie algebra is
faithful\supcite{Sat,Ise}) in the irreducible decomposition
(\ref{eq:svnsvn}), which holds for $q=1$.
As we have studied in the preceding section,
(\ref{eq:RTT})--(\ref{eq:TFC}) with $\repr{i}{j}$ in place of
$\tgn{i}{j}$ are right equations in $U(\gee_2)\dual$.  Hence,
considering the pairing between each one of these equations
and $x:=e^X$ for $X\in\gee_2\subset U(\gee_2)$, and using the equations
$\Delta(x)=x\otimes x$ and $\;\epsilon\,(x)=1$ satisfied at $q=1$,
we easily find $(\repr{i}{j}(x))\in G$.
Such elements $(\repr{i}{j}(x))$ generate a connected Lie group
with the Lie algebra $\gee_2$ because $\rep{7}$ is faithful.
Hence $\Aut \Cayley \subset G$ follows.

So we arrive at $G = \Aut\Cayley$.
\end{proof}
\mytuning{\pagebreak}

The proposition implies that the Hopf algebra $A(q)$ is considered
as a quantization of the $\Gtwo$ type Lie group $\Aut\Cayley$; at $q=1$,
$\tgn{i}{j}$ can be viewed as the tautological mappings
$(\crdn{i}{j}) \mapsto \crdn{i}{j}$, which generate
$\Fun(\Gtwo) := \Fun(\Aut \Cayley)$.
Henceforth $A(q)$ is denoted by $\Fun_q(\Gtwo)$.

It is possible in practice to determine the explicit form of the
structure constants of $\Fun_q(\Gtwo)$;
the $R$-matrix $(\Rmat{i}{j}{k}{l})$ is already known,\supcite{Res}
while the other constants
$\CGg{i}{j}$, $\CGgc{i}{j}$, $\CGf{i}{j}{k}$, $\CGfc{i}{j}{k}$ are
rather easily determined ($\CGf{i}{j}{k}$ in particular is also
already given\supcite{Res} explicitly).
With the help of Reduce --- a computer software for mathematical-formula
processing --- the author also computed the structure constants of
$\Fun_q(\Gtwo)$ with the convention described in Sec.~\ref{sec-nc} with
\begin{eqnarray}
& a_{11} = 2,\;\; a_{12} = -1,\;\; a_{21} = -3,\;\; a_{22} = 2,
\spc\spc q_1 = q^{3},\;\; q_2 = q &
\end{eqnarray}
and obtained the following decomposition:
\begin{eqnarray}
&\widehat{R}:= PR
 = q^{-12}\,P_{\rep{1}} - q^{-6}\,P_{\rep{7}}
 - P_{\rep{14}} + q^{2}\,P_{\rep{27}},&
\end{eqnarray}
where $P=(P^{ij}{}_{kl}):=(\Kdel{i}{l}\Kdel{j}{k})$ stands for the
permutation matrix and $P_{\rep{n}}$ stands for the projection matrix
that extracts the irreducible component isomorphic to $\rep{n}$ from the
tensor product representation $\rep{7}\otimes\rep{7}$.  That the matrix
$\widehat{R}$ has a decomposition into the projections is an immediate
consequence of (\ref{eq:intwn}) and (\ref{eq:svnsvn}).
The author determined the eignvalues by virtue of
(\ref{eq:RRg})--(\ref{eq:fcRR}) with using explicit form of
$\CGg{i}{j}$, $\CGf{i}{j}{k}$ and $P_{\rep{n}}$; much more
systematic way to derive the decomposition had been given by
Reshetikhin\supcite{Res} (see Ref.~\incite{Gou} also).

Before giving the explicit form of the structure constants, we think it
in order to clarifying the basis of $\rep{7}$ adopted in the calculation;
we use the basis that gives the following representation matrices:
\begin{eqnarray*}
\bigl(\repr{i}{j}(K_1)\bigr)
 \rop{=} \diag(\,1,\,\,q^{3},\:q^{-3},\:\,1,\,\,q^{3},\:q^{-3},\,\,1\,), \\
\bigl(\repr{i}{j}(K_2)\bigr)
 \rop{=} \diag(\,q,\,q^{-1},\;q^{2},\:\,1,\,q^{-2},\;q,\,q^{-1}), \\
\bigl(\repr{i}{j}(X_1^{+})\bigr)
 \rop{=} \bigl( 0 \bigr) \oplus
 \left( \begin{array}{cc} 0 & 1 \\ 0 & 0 \end{array} \right)
 \oplus \bigl( 0 \bigr) \oplus
 \left( \begin{array}{cc} 0 & 1 \\ 0 & 0 \end{array} \right)
 \oplus \bigl( 0 \bigr),  \\
\bigl(\repr{i}{j}(X_2^{+})\bigr)
 \rop{=}
 \left( \begin{array}{cc} 0 & 1 \\ 0 & 0 \end{array} \right)
 \oplus \left( \begin{array}{ccc}
   0 & [2] & 0 \\ 0 & 0 & 1 \\ 0 & 0 & 0
   \end{array} \right) \oplus
 \left( \begin{array}{cc} 0 & 1 \\ 0 & 0 \end{array} \right), \\
\bigl(\repr{i}{j}(X_1^{-})\bigr)
 \rop{=} \bigl( 0 \bigr) \oplus
 \left( \begin{array}{cc} 0 & 0 \\ 1 & 0 \end{array} \right)
 \oplus \bigl( 0 \bigr) \oplus
 \left( \begin{array}{cc} 0 & 0 \\ 1 & 0 \end{array} \right)
 \oplus \bigl( 0 \bigr),  \\
\bigl(\repr{i}{j}(X_2^{-})\bigr)
 \rop{=}
 \left( \begin{array}{cc} 0 & 0 \\ 1 & 0 \end{array} \right)
 \oplus \left( \begin{array}{ccc}
   0 & 0 & 0 \\ 1 & 0 & 0 \\ 0 & [2] & 0
   \end{array} \right) \oplus
 \left( \begin{array}{cc} 0 & 0 \\ 1 & 0 \end{array} \right),
\end{eqnarray*}
where $[2] := q + q^{-1}$ and $i$ is a row index whereas $j$ is a column
one.  In this setting, all the nonzero components among the structure
constants of $\Fun_q(\Gtwo)$ are as follows:
\mytuning{\newpage\noindent}There are 112 nonzero entries of the $R$-matrix
\begin{eqnarray*}
&&
  \Rmx{1}{1}{1}{1} = \Rmx{2}{2}{2}{2} = \Rmx{3}{3}{3}{3} = \Rmx{5}{5}{5}{5}
= \Rmx{6}{6}{6}{6} = \Rmx{7}{7}{7}{7} = q^{2} ,\\
&&
  \Rmx{1}{2}{1}{2} = \Rmx{1}{3}{1}{3} = \Rmx{2}{1}{2}{1} = \Rmx{2}{5}{2}{5}
= \Rmx{3}{1}{3}{1} = \Rmx{3}{6}{3}{6} = \Rmx{5}{2}{5}{2} = \Rmx{5}{7}{5}{7}
  \\ && \;\;       = \Rmx{6}{3}{6}{3} = \Rmx{6}{7}{6}{7} = \Rmx{7}{5}{7}{5}
= \Rmx{7}{6}{7}{6} = q ,\\
&&
  \Rmx{1}{4}{1}{4} = \Rmx{2}{4}{2}{4} = \Rmx{3}{4}{3}{4} = \Rmx{4}{1}{4}{1}
= \Rmx{4}{2}{4}{2} = \Rmx{4}{3}{4}{3} = \Rmx{4}{4}{4}{4} = \Rmx{4}{5}{4}{5}
  \\ && \;\;       = \Rmx{4}{6}{4}{6} = \Rmx{4}{7}{4}{7} = \Rmx{5}{4}{5}{4}
= \Rmx{6}{4}{6}{4} = \Rmx{7}{4}{7}{4} = 1 ,\\
&&
  \Rmx{1}{5}{1}{5} = \Rmx{1}{6}{1}{6} = \Rmx{2}{3}{2}{3} = \Rmx{2}{7}{2}{7}
= \Rmx{3}{2}{3}{2} = \Rmx{3}{7}{3}{7} = \Rmx{5}{1}{5}{1} = \Rmx{5}{6}{5}{6}
  \\ && \;\;       = \Rmx{6}{1}{6}{1} = \Rmx{6}{5}{6}{5} = \Rmx{7}{2}{7}{2}
= \Rmx{7}{3}{7}{3} = q^{-1} ,\\
&&
  \Rmx{1}{7}{1}{7} = \Rmx{2}{6}{2}{6} = \Rmx{3}{5}{3}{5} = \Rmx{5}{3}{5}{3}
= \Rmx{6}{2}{6}{2} = \Rmx{7}{1}{7}{1} = q^{-2} ,\\
&&
  \Rmx{2}{1}{1}{2} = \Rmx{3}{1}{1}{3} = \Rmx{5}{2}{2}{5} = \Rmx{6}{3}{3}{6}
= \Rmx{7}{5}{5}{7} = \Rmx{7}{6}{6}{7} = q^{2}-1 ,\\
&&
  \Rmx{2}{3}{1}{4} = \Rmx{5}{1}{4}{2} = \Rmx{5}{6}{4}{7} = \Rmx{6}{1}{4}{3}
= \Rmx{7}{2}{5}{4} = \Rmx{7}{3}{6}{4} = q-q^{-3} ,\\
&&
  \Rmx{2}{4}{1}{5} = \Rmx{3}{4}{1}{6} = \Rmx{4}{1}{3}{2} = \Rmx{4}{4}{3}{5}
= \Rmx{4}{5}{2}{7} = \Rmx{4}{6}{3}{7} = \Rmx{7}{4}{6}{5} = q-q^{-1} ,\\
&&
  \Rmx{2}{6}{1}{7} = \Rmx{7}{1}{6}{2} = q^{-1}-q^{-3} ,\\
&&
  \Rmx{3}{2}{1}{4} = \Rmx{6}{5}{4}{7} = -q^{-2}+q^{-6} ,\\
&&
  \Rmx{3}{2}{2}{3} = \Rmx{6}{5}{5}{6} = q^{2}-q^{-4} ,\\
&&
  \Rmx{3}{5}{1}{7} = \Rmx{7}{1}{5}{3} = -q^{-4}+q^{-6} ,\\
&&
  \Rmx{3}{5}{2}{6} = \Rmx{6}{2}{5}{3} = q-q^{-5} ,\\
&&
  \Rmx{4}{1}{1}{4} = \Rmx{7}{4}{4}{7} = q^{2}-1+q^{-4}-q^{-6} ,\\
&&
  \Rmx{4}{1}{2}{3} = \Rmx{4}{4}{2}{6} = \Rmx{7}{4}{5}{6} = -q^{-2}+q^{-4} ,\\
&&
  \Rmx{4}{2}{1}{5} = \Rmx{4}{3}{1}{6} = \Rmx{5}{4}{2}{7}
= \Rmx{6}{4}{3}{7} = -q^{-1}+q^{-3} ,\\
&&
  \Rmx{4}{2}{2}{4} = \Rmx{4}{3}{3}{4} = \Rmx{5}{4}{4}{5} = \Rmx{6}{4}{4}{6}
=  q^{2}-q^{-2} ,\\
&&
  \Rmx{4}{4}{1}{7} = q-2q^{-1}+2q^{-3}-q^{-5} ,\\
&&
  \Rmx{5}{1}{1}{5} = \Rmx{6}{1}{1}{6} = \Rmx{7}{2}{2}{7} = \Rmx{7}{3}{3}{7}
=  q^{2}-1+q^{-2}-q^{-6} ,\\
&&
  \Rmx{5}{1}{2}{4} = \Rmx{6}{1}{3}{4} = \Rmx{7}{2}{4}{5} = \Rmx{7}{3}{4}{6}
=  -q^{-1}+q^{-5} ,\\
&&
  \Rmx{5}{3}{1}{7} = \Rmx{7}{1}{3}{5} = -1+q^{-4}-q^{-6}+q^{-8} ,\\
&&
  \Rmx{5}{3}{2}{6} = \Rmx{6}{2}{3}{5} = q^{-5}-q^{-7} ,\\
&&
  \Rmx{5}{3}{3}{5} = q^{2}-1-q^{-2}+q^{-4} ,\\
&&
  \Rmx{5}{3}{4}{4} = q+q^{-1}-q^{-3}-q^{-5} ,\\
&&
  \Rmx{6}{2}{1}{7} = \Rmx{7}{1}{2}{6} = q^{-3}-q^{-7}+q^{-9}-q^{-11} ,\\
&&
  \Rmx{6}{2}{2}{6} = q^{2}-1-q^{-8}+q^{-10} ,\\
&&
  \Rmx{6}{2}{4}{4} = -q^{-2}-q^{-4}+q^{-6}+q^{-8} ,\\
&&
  \Rmx{7}{1}{1}{7} = q^{2}-1+q^{-2}-q^{-4}-q^{-6}+q^{-8}-q^{-10}+q^{-12} ,\\
&&
  \Rmx{7}{1}{4}{4} = q-q^{-3}+q^{-5}-q^{-9} .
\end{eqnarray*}
\mytuning{\newpage\noindent}There are $2\,\times\,(7+31)$ nonzero structure
constants of Fun$_q($G$_2)$ other than the entries of the $R$-matrix
\begin{eqnarray*}
&& \SCg{1}{7} = q^{5} ,\wdsp
   \SCg{2}{6} = -q^{4} ,\wdsp
   \SCg{3}{5} = q ,\wdsp
   \SCg{4}{4} = {-1}/(1+q^{-2}) ,\\
&& \SCg{5}{3} = q^{-1} ,\wdsp
   \SCg{6}{2} = -q^{-4} ,\wdsp
   \SCg{7}{1} = q^{-5} ,\\
\tallsp
&&
  \SCf{1}{4}{1} = \SCf{4}{7}{7} = -q^{3} ,\\
&&
  \SCf{1}{5}{2} = \SCf{1}{6}{3} = \SCf{2}{6}{4} = \SCf{2}{7}{5}
= \SCf{3}{7}{6} = -q^{3}-q ,\\
&&
  \SCf{1}{7}{4} = -q^{2}-1 ,\;\;\wdsp\wdsp
  \SCf{2}{3}{1} = \SCf{5}{6}{7} = q^{2}+1 ,\\
&&
  \SCf{2}{4}{2} = \SCf{3}{4}{3} = \SCf{4}{5}{5} = \SCf{4}{6}{6} = q ,\\
&&
  \SCf{3}{2}{1} = \SCf{6}{5}{7} = -q^{-1}-q^{-3} ,\\
&&
  \SCf{3}{5}{4} = 1+q^{-2} ,\;\;\;\wdsp\wdsp
  \SCf{4}{1}{1} = \SCf{7}{4}{7} = q^{-3} ,\\
&&
  \SCf{4}{2}{2} = \SCf{4}{3}{3} = \SCf{5}{4}{5} = \SCf{6}{4}{6} = -q^{-1} ,\\
&&
  \SCf{4}{4}{4} = q-q^{-1} ,\\
&&
  \SCf{5}{1}{2} = \SCf{6}{1}{3} = \SCf{7}{1}{4} = \SCf{7}{2}{5}
= \SCf{7}{3}{6} = q^{-2}+q^{-4} ,\\
&&
  \SCf{5}{3}{4} = -1-q^{-2} ,\wdsp\wdsp
  \SCf{6}{2}{4} = q^{-3}+q^{-5} ,\\
\tallsp
&& \Sgc{1}{7} = q^{-5} ,\wdsp
   \Sgc{2}{6} = -q^{-4} ,\wdsp
   \Sgc{3}{5} = q^{-1} ,\wdsp
   \Sgc{4}{4} = -1-q^{-2} ,\\
&& \Sgc{5}{3} = q ,\wdsp
   \Sgc{6}{2} = -q^{4} ,\wdsp
   \Sgc{7}{1} = q^{5} ,\\
\tallsp
&&
  \Sfc{1}{1}{4} = \Sfc{7}{4}{7} = -q^{-3}-q^{-5} ,\\
&&
  \Sfc{1}{2}{3} = \Sfc{2}{2}{4} = \Sfc{3}{3}{4} = \Sfc{5}{4}{5}
= \Sfc{6}{4}{6} = \Sfc{7}{5}{6} = q^{-1}+q^{-3} ,\\
&&
  \Sfc{1}{3}{2} = \Sfc{7}{6}{5} = -q^{2}-1 ,\\
&&
  \Sfc{1}{4}{1} = \Sfc{2}{5}{1} = \Sfc{5}{7}{2} = \Sfc{6}{7}{3}
= \Sfc{7}{7}{4} = q^{3}+q ,\\
&&
  \Sfc{2}{1}{5} = \Sfc{3}{1}{6} = \Sfc{5}{2}{7} = \Sfc{6}{3}{7}
=  -q^{-2}-q^{-4} ,\\
&&
  \Sfc{2}{4}{2} = \Sfc{3}{4}{3} = \Sfc{5}{5}{4} = \Sfc{6}{6}{4}
=  -q-q^{-1} ,\\
&&
  \Sfc{3}{6}{1} = q^{3}+q ,\;\;\;\;\;\wdsp\wdsp
  \Sfc{4}{1}{7} = -q^{-2} ,\\
&&
  \Sfc{4}{2}{6} = -q^{-3} ,\;\;\;\;\;\;\wdsp\wdsp
  \Sfc{4}{3}{5} = 1 ,\\
&&
  \Sfc{4}{4}{4} = -q+q^{-3} ,\wdsp\wdsp
  \Sfc{4}{5}{3} = -1 ,\\
&&
  \Sfc{4}{6}{2} = q^{3} ,\;\;\;\;\;\;\;\;\;\;\;\wdsp\wdsp
  \Sfc{4}{7}{1} = q^{2} ,
\end{eqnarray*}
which satisfy $\CGfc{i}{k}{l}\,\CGf{l}{k}{j}
 = -(1+q^{-2})(q^{4}+q^{-4})(q^{2}+1+q^{-2})\:\delta^{i}{}_{j}
$ \,({\it cf.} (\ref{eq:FFsId})).

The explicit form of the $R$-matrix exhibits
\begin{eqnarray*}
&7\,(i-1)+j \:<\: 7\,(k-1)+l&
 \spc \Rightarrow \spc\; \Rmat{i}{j}{k}{l} = 0,\\
&7\,(j-1)+i \:>\: 7\,(\,l-1)+k&
 \spc \Rightarrow \spc\; \Rmat{i}{j}{k}{l} = 0.
\end{eqnarray*}
These with (\ref{eq:ldef}) state that $L^{(-)}:=(\lgn{-}{i}{j})$ is lower
triangular whereas $L^{(+)}:=(\lgn{+}{i}{j})$ is upper triangular in this
setting.

\mytuning{\newpage}
\section*{acknowledgements}
The author is indebted to Prof.~R.~Sasaki
for important information and helpful advices.
The author is grateful to Dr.~Y.~Yasui and Dr.~M.~Takama for
giving him a good opportunity to study quantum groups and
for helpful conversations on this work.  The author is also grateful
to M.~Miyajima for his help in computing the constants.

\end{document}